\documentclass[12pt]{article}
\usepackage{graphics}
\begin{document}

\title{Entangling two mode thermal fields through quantum erasing}
\author{Shang-Bin Li$^{1,2}$ and Jing-Bo Xu$^{1,2}$\\
$^{1}$Chinese Center of Advanced Science and Technology (World Labor-\\
atory), P.O.Box 8730, Beijing, People's Republic of China;\\
$^{2}$Zhejiang Institute of Modern Physics and Department of Physics,\\
Zhejiang University, Hangzhou 310027, People's Republic of
China\thanks{Mailing address}}
\date{}
\maketitle

\begin{abstract}
{\normalsize We investigate a possible scheme for entangling two
mode thermal fields through the quantum erasing process, in which
an atom is coupled with two mode fields via the interaction
governed by the two-mode two-photon Jaynes-Cummings model. The
influence of phase decoherence on the entanglement of two mode
fields is discussed. It is found that quantum erasing process can
transfer part of entanglement between the atom and fields to two
mode fields initially in the thermal states. The entanglement
achieved by fields heavily depends on their initial temperature
and the detuning. The entanglement of stationary state is
also investigated.\\

PACS numbers: 03.67.Mn, 03.67.-a, 03.65.Fd}
\end{abstract}
\newpage
\section * {I. INTRODUCTION}

Entanglement, an important resource for quantum information
processing [1], is one of the most prominent nonclassical
properties in quantum theory. Entanglement can exhibit a nonlocal
correlation between quantum systems that have no classical
interpretation. Recently, much attention has been focused on the
entanglement in bipartite or multipartite systems in which the
subsystems are initially in thermal equilibrium [2,3,4,5]. Arnesen
et al. have shown that a natural entanglement arises in Heisenberg
spin chain in thermal equilibrium, and the entanglement can be
improved by increasing the temperature [2]. Instead of attempting
to shield the system from the environmental noise, Plenio and
Huelge [3] use white noise to play a constructive role and
generate the controllable entanglement by incoherent sources.
Similar work on this aspect has also been considered by other
authors [4,5]. However, very little attention has been paid to the
study of entangling two mode thermal fields. In this paper, we
investigate a possible scheme for entangling two mode thermal
fields through the quantum erasing process, in which an atom is
coupled with two mode fields via the interaction governed by the
two-mode two-photon Jaynes-Cummings model. The influence of phase
decoherence on the entanglement of two mode fields is discussed.
It is found that quantum erasing process can transfer part of
entanglement between the atom and fields to two mode fields
initially in the thermal states. The entanglement achieved by
fields heavily depends on their initial temperature and the
detuning. The term "quantum eraser" [6] was invented to describe
the loss or gain of interference or, more generally quantum
information, in a subensemble, based on the measurement outcomes
of two complementary observables. It was reported that the
implementation of two- and three-spin quantum eraser using nuclear
magnetic resonance, and shown that quantum erasers provide a means
of manipulating quantum entanglement [7]. The quantum erasing
process discussed in this paper is implemented by measuring the
polarizing vector of a two-level atom coupling with two mode
quantum fields. The project measurement of an atom
has been extensively studied both in the theoretical and experimental aspects.\\
This paper is organized as follows. In Sec.II, we study the system
in which an atom is coupled with two mode fields via the
interaction governed by the two-mode two-photon Jaynes-Cummings
model by making use of the dynamical algebraical method [8,9] and
find the exact solution of the master equation for the system with
phase decoherence. Based on the exact solution, we then propose a
possible way to entangle two mode thermal fields through the
quantum erasing process, which is realized by measuring the atom.
In Sec.III, we use the log-negativity to characterize the
entanglement between two mode fields. It is shown that quantum
erasing process can transfer part of entanglement between the atom
and fields to two mode fields initially in the thermal states. The
entanglement achieved by fields heavily depends on their initial
temperature and the detuning. A conclusion is given in Sec.IV.\\

\section * {II. SOLUTION OF AN ATOM COUPLES TO TWO THERMAL FIELDS WITH PHASE DECOHERENCE}
We consider the two-mode two-photon Jaynes-Cummings model [10].
The Hamiltonian for the model can be described by ($\hbar=1$),
$$
H=\omega_1a^{\dagger}_1a_1+\omega_2a^{\dagger}_2a_2+\frac{\omega}{2}\sigma_z+g(a_1a_2\sigma_{+}+a^{\dagger}_1a^{\dagger}_2\sigma_{-}),
\eqno{(1)}
$$
where $\sigma_z$ and $\sigma_{\pm}$ are the atomic spin flip
operators characterizing the effective two-level atom with
transition frequency $\omega$ and $a_1$ ($a_2$), $a^{\dagger}_1$
($a^{\dagger}_2$) are annihilation and creation operators of the
first (second) mode light field of frequencies $\omega_1$
($\omega_2$) respectively. The Hamiltonian (1) ignores Stark
shifts and the parameter $g$ is the atom-field coupling
constant.\\

It is easy to see that there exist two constants of motion in the
Hamiltonian (1),
$$
K_1=a^{\dagger}_1a_1+\frac{1+\sigma_z}{2},~~~K_2=a^{\dagger}_2a_2+\frac{1+\sigma_z}{2},
\eqno{(2)}
$$
which commute not only with Hamiltonian but also with operators
$a_1a_2\sigma_{+}$ and $a^{\dagger}_1a^{\dagger}_2\sigma_{-}$. We
can introduce the following operators
$$
S_0=\frac{\sigma_z}{2},~~~S_+=\frac{a_1a_2\sigma_{+}}{\sqrt{K_1K_2}},~~~S_-=\frac{a^{\dagger}_1a^{\dagger}_2\sigma_{-}}{\sqrt{K_1K_2}}.
\eqno{(3)}
$$
The operators $S_{\pm}$ and $S_0$ satisfy the following
commutation relations
$$
[S_0,S_{\pm}]=\pm{S}_{\pm},~~~[S_+,S_-]=2S_0,
\eqno{(4)}
$$
where $S_{\pm}$ and $S_0$ are the generators of the su(2) algebra.
In terms of the su(2) generators, we can rewrite the Hamiltonian
(1) as
$$
H=\omega_1(K_1-\frac{1}{2})+\omega_2(K_2-\frac{1}{2})+\Delta{S}_0+g\sqrt{K_1K_2}(S_++S_-),
\eqno{(5)}
$$
where $\Delta=\omega-\omega_1-\omega_2$. With the help of the
su(2) dynamical algebraic structure, we can diagonalize the
Hamiltonian (5) by introducing a unitary transformation
$$
U=\exp[\frac{\theta(K_1,K_2)}{2}(S_+-S_-)]
\eqno{(6)}
$$
with $\theta(K_1,K_2)=\arctan(2g\sqrt{K_1K_2}/\Delta)$, and get
transformed Hamiltonian
$$
H^{\prime}=UHU^{\dagger}=\omega_1(K_1-\frac{1}{2})+\omega_2(K_2-\frac{1}{2})+2\Omega(K_1,K_2)S_0,
\eqno{(7)}
$$
where $\Omega(K_1,K_2)=\sqrt{\Delta^2/4+g^2K_1K_2}$.

In this paper, we consider the phase decoherence mechanism only.
In this situation, the master equation governing the time
evolution for the system under the Markovian approximation is
given by [11]
$$
\frac{d\rho}{dt}=-i[H,\rho]-\frac{\gamma}{2}[H,[H,\rho]],
\eqno{(8)}
$$
where $\gamma$ is the phase decoherence coefficient. Noted that
the equation with the similar form has been proposed to describing
the intrinsic decoherence [12]. The formal solution of the master
equation (8) can be expressed as follows [13],
$$
\rho(t)=\sum^{\infty}_{k=0}\frac{(\gamma{t})^{k}}{k!}M^{k}(t)\rho(0)M^{\dagger{k}}(t),
\eqno{(9)}
$$
where $\rho(0)$ is the density operators of the initial atom-field
system and $M^{k}(t)$ is defined by
$$
M^k(t)=H^k\exp(-iHt)\exp(-\frac{\gamma{t}}{2}H^2). \eqno{(10)}
$$
By means of the SU(2) dynamical algebraic structure, we obtain the
explicit expression for the operator $M^k$
$$
M^k(t)=U^{\dagger}H^{\prime{k}}\exp(-iH^{\prime}t)\exp(-\frac{\gamma{t}}{2}H^{\prime2})U
$$
$$
~~~=\frac{1}{2}[\hat{f}^k_{+}\exp(-i\hat{f}_{+}t)\exp(-\frac{\gamma{t}\hat{f}^2_{+}}{2})+\hat{f}^k_{-}\exp(-i\hat{f}_{-}t)\exp(-\frac{\gamma{t}\hat{f}^2_{-}}{2})]
$$
$$
~~~+\frac{1}{2}[\hat{f}^k_{+}\exp(-i\hat{f}_{+}t)\exp(-\frac{\gamma{t}\hat{f}^2_{+}}{2})-\hat{f}^k_{-}\exp(-i\hat{f}_{-}t)\exp(-\frac{\gamma{t}\hat{f}^2_{-}}{2})][\frac{\Delta\sigma_z}{2\Omega(K_1,K_2)}+\frac{g(a_1a_2\sigma_++a^{\dagger}_1a^{\dagger}_2\sigma_-)}{\Omega(K_1,K_2)}],
\eqno{(11)}
$$
where
$\hat{f}_{\pm}=\omega_1(K_1-1/2)+\omega_2(K_2-1/2)\pm\Omega(K_1,K_2)$.
Firstly, we assume that the cavity fields are initially in
two-mode Fock states $|n_1n_2\rangle$, and the atom is in the
excited state $|e\rangle$. The time evolution of $\rho(t)$ can be
written as follows,
$$
\rho(t)=\frac{1}{4}[2+\frac{\Delta^2}{2\Omega^2_{n_1,n_2}}+(2-\frac{\Delta^2}{2\Omega^2_{n_1,n_2}})e^{-2\gamma{t}\Omega^2_{n_1,n_2}}\cos2\Omega_{n_1,n_2}t]|n_1,n_2\rangle\langle{n_1},n_2|
\otimes|e\rangle\langle{e}|
$$
$$
~~~+\frac{1}{4}\frac{g^2(n_1+1)(n_2+1)}{\Omega^2_{n_1,n_2}}[2-2e^{-2\gamma{t}\Omega^2_{n_1,n_2}}\cos2\Omega_{n_1,n_2}t]|n_1+1,n_2+1\rangle\langle{n_1+1},n_2+1|\otimes|g\rangle\langle{g}|
$$
$$
~~~+\frac{g\sqrt{(n_1+1)(n_2+1)}}{4\Omega_{n_1,n_2}}\{\frac{\Delta}{\Omega_{n_1,n_2}}[1-e^{-2\gamma{t}\Omega^2_{n_1,n_2}}\cos2\Omega_{n_1,n_2}t]
$$
$$
~~~+2ie^{-2\gamma{t}\Omega^2_{n_1,n_2}}\sin2\Omega_{n_1,n_2}t\}|n_1,n_2\rangle\langle{n_1+1},n_2+1|\otimes|e\rangle\langle{g}|
$$
$$
~~~+\frac{g\sqrt{(n_1+1)(n_2+1)}}{4\Omega_{n_1,n_2}}\{\frac{\Delta}{\Omega_{n_1,n_2}}[1-e^{-2\gamma{t}\Omega^2_{n_1,n_2}}\cos2\Omega_{n_1,n_2}t]
$$
$$
~~~-2ie^{-2\gamma{t}\Omega^2_{n_1,n_2}}\sin2\Omega_{n_1,n_2}t\}|n_1+1,n_2+1\rangle\langle{n_1},n_2|\otimes|g\rangle\langle{e}|,
\eqno{(12)}
$$
where $\Omega_{n_1,n_2}=\sqrt{\Delta^2/4+g^2(n_1+1)(n_2+1)}$. In
the basis $\{|1,1\rangle\equiv|n_1,n_2\rangle\otimes|e\rangle,
|0,1\rangle\equiv|n_1+1,n_2+1\rangle\otimes|e\rangle,|1,0\rangle\equiv|n_1,n_2\rangle\otimes|g\rangle,|0,0\rangle\equiv|n_1+1,n_2+1\rangle\otimes|g\rangle,\}$,
$\rho(t)$ can be regarded as a two qubit mixed state. Then, a
quantum erasing is applied to this system by making a project
measurement of the atom on the basis
$\{\cos\frac{\theta}{2}|e\rangle+e^{i\phi}\sin\frac{\theta}{2}|g\rangle,\cos\frac{\theta}{2}|g\rangle-e^{-i\phi}\sin\frac{\theta}{2}|e\rangle\}$.
It is easy to verify that two fields will get the same amount of
entanglement corresponding to two different measurement outcomes
if the value of $\theta$ is $\pi/2$. In this case, both the
probabilities of two projection measurement results are
$\frac{1}{2}$. So we can only consider the entanglement of one of
the projection results instead of average entanglement between the
two fields after the measurement. If the measurement projects the
state of the atom onto
$\cos\frac{\theta}{2}|e\rangle+e^{i\phi}\sin\frac{\theta}{2}|g\rangle$,
the residual state of two mode fields is expressed by
(unnormalized)
$$
\rho_f(n_1,n_2,t)=\frac{1}{4}\cos^2\frac{\theta}{2}[2+\frac{\Delta^2}{2\Omega^2_{n_1,n_2}}+(2-\frac{\Delta^2}{2\Omega^2_{n_1,n_2}})e^{-2\gamma{t}\Omega^2_{n_1,n_2}}\cos2\Omega_{n_1,n_2}t]||n_1,n_2\rangle\langle{n_1,n_2}|
$$
$$
+\frac{1}{4}\sin^2\frac{\theta}{2}\frac{g^2(n_1+1)(n_2+1)}{\Omega^2_{n_1,n_2}}[2-2e^{-2\gamma{t}\Omega^2_{n_1,n_2}}\cos2\Omega_{n_1,n_2}t]||n_1+1,n_2+1\rangle\langle{n_1+1,n_2+1}|
$$
$$
+\frac{1}{8}\sin\theta{e}^{i\phi}\frac{g\sqrt{(n_1+1)(n_2+1)}}{\Omega_{n_1,n_2}}\{\frac{\Delta}{\Omega_{n_1,n_2}}[1-e^{-2\gamma{t}\Omega^2_{n_1,n_2}}\cos2\Omega_{n_1,n_2}t]
$$
$$
~~~~~~+2ie^{-2\gamma{t}\Omega^2_{n_1,n_2}}\sin2\Omega_{n_1,n_2}t\}||n_1,n_2\rangle\langle{n_1+1,n_2+1}|
$$
$$
+\frac{1}{8}\sin\theta{e}^{-i\phi}\frac{g\sqrt{(n_1+1)(n_2+1)}}{\Omega_{n_1,n_2}}\{\frac{\Delta}{\Omega_{n_1,n_2}}[1-e^{-2\gamma{t}\Omega^2_{n_1,n_2}}\cos2\Omega_{n_1,n_2}t]
$$
$$
~~~~~~-2ie^{-2\gamma{t}\Omega^2_{n_1,n_2}}\sin2\Omega_{n_1,n_2}t\}||n_1+1,n_2+1\rangle\langle{n_1,n_2}|
\eqno{(13)}
$$
For the initial two mode thermal fields, the output state of two
fields is replaced by
$$
\rho_f(t)={\mathcal{N}}\sum^{\infty}_{n_1,n_2=0}\frac{\bar{m}^{n_1}_1\bar{m}^{n_2}_2}{(1+\bar{m}_1)^{n_1+1}(1+\bar{m}_2)^{n_2+1}}\rho_f(n_1,n_2,t),
\eqno{(14)}
$$
where
${\mathcal{N}}=\{\sum^{\infty}_{n_1,n_2=0}\frac{\bar{m}^{n_1}_1\bar{m}^{n_2}_2}{(1+\bar{m}_1)^{n_1+1}(1+\bar{m}_2)^{n_2+1}}{\mathrm{Tr}}[\rho_f(n_1,n_2,t)]\}^{-1}$
is the normalization constant, and
$\bar{m}_i=[\exp(\beta_i\omega_i)-1]^{-1}$ ($i=1,2$) is the mean
photon number of the $i$th mode thermal field at the inverse
temperature $\beta_i$.
\section * {III. THE LOG-NEGATIVITY OF TWO MODE FIELDS}

In order to quantify the degree of entanglement, we adopt the
log-negativity $N(\rho)$ to calculate the entanglement between two
mode fields, which is defined as [14]
$$
N(\rho)=\log_2\|\rho^{\Gamma}\|, \eqno{(15)}
$$
where $\rho^{\Gamma}$ is the partial transpose of $\rho$ and $\|\rho^{\Gamma}\|$ denotes the
trace norm of $\rho^{\Gamma}$, which is the sum of the singular values of $\rho^{\Gamma}$.\\

For the unnormalized density operator in Eq.(13), it is easy to
derive its stationary log-negativity which is given by
$\log_2[1+2|\sin\theta\frac{g\Delta\sqrt{(n_1+1)(n_2+1)}}{4\Omega^2_{n_1,n_2}+\Delta^2\cos\theta}|]$.
For simplicity, we will set the value of $\theta$ as
$\frac{\pi}{2}$ throughout the following calculation. First of
all, one important fact should be pointed that the entanglement
between two initial thermal fields can not arise if the quantum
erasing processing is not applied and the degree of freedom of the
atom is simply traced. The entanglement between the fields is
partly transferred from the entanglement between the atom and the
fields through the quantum erasing. So, all of the following
discussions concerning the entanglement between two fields at any
time $t$ are based on the presumption that a projection
measurement is just acted on the atom at the time $t^{-}$. In
Fig.1, the stationary state log-negativity $N(\rho_f)$ of the
density operator $\rho_f(\infty)$ is plotted as a function of the
mean photon number $\bar{m}_1=\bar{m}_2=\alpha$ of initial thermal
fields and the detuning $\Delta$. Fig.1 shows that, in the
resonant case $\Delta=0$, there is not any entanglement in the
stationary state. In the off-resonant case, the entanglement
decreases with $\alpha$, and eventually disappears as the value of
$\alpha$ goes beyond a threshold value which is dependent on the
detuning. A natural question will arise how the entanglement
behaves when one mode is initially in the vacuum state and the
other mode is in thermal state. In Ref.[15], the authors indicated
that the subsystem purity can enforce the entanglement. So, it is
easy to understand the result displayed in Fig.2, where the
stationary state log-negativity $N(\rho_f)$ of the density
operator $\rho_f(\infty)$ is plotted as a function of the mean
photon number $\bar{m}_2=\alpha$ of initial thermal field of the
second mode and the detuning $\Delta$ with $\bar{m}_1=0$, i.e.,
the first mode is initially in a vacuum state. We can find that
the entanglement always exists for any high temperature of the
second mode in the off-resonant situation. One may also conjecture
that the stationary state entanglement can increase with the
difference of the mean photon numbers of two thermal fields as the
value of $\bar{m}_1+\bar{m}_2$ is fixed. This seems to be true,
and can be seen from the Fig.3, in which we depict the stationary
entanglement as a function of $\bar{m}_1$ and $\bar{m}_2$ with
$g=0.5$ and $\Delta=1$. The stationary entanglement always
increases with the value of $|\bar{m}_1-\bar{m}_2|$ along any line
characterized by $\bar{m}_1+\bar{m}_2={\mathrm{const.}}$. We
conjecture this phenomenon exists in a wide class of systems,
including the thermal modes in different thermal reserviors
effectively coupled by the qubits in a quantum register.\\
\begin{figure}
\centering
\includegraphics{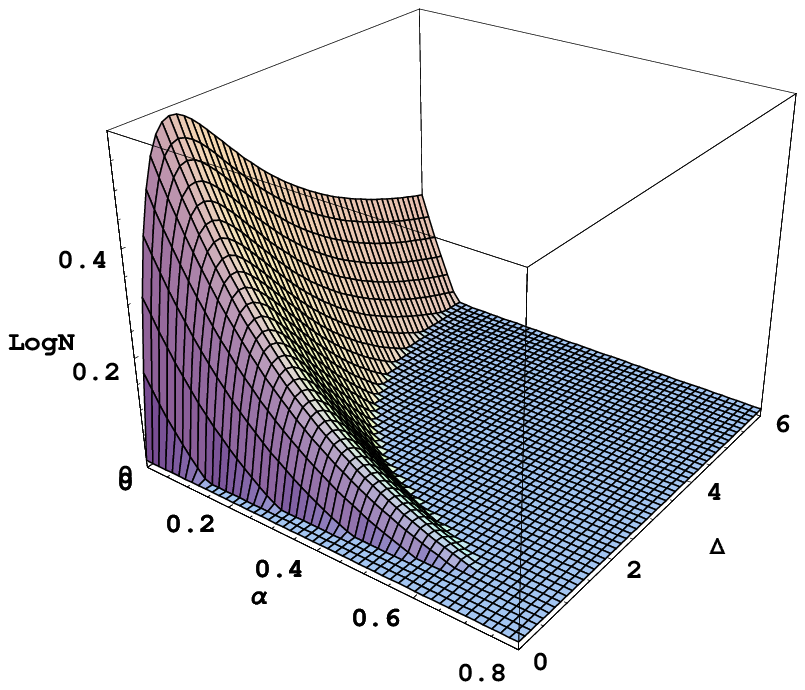}
\caption{The stationary state log-negativity $N(\rho_f)$ of the
density operator $\rho_f(\infty)$ is plotted as a function of the
mean photon number $\bar{m}_1=\bar{m}_2=\alpha$ of initial thermal
fields and the detuning $\Delta$ with $g=0.5$ and
$\theta=\frac{\pi}{2}$. \label{Fig1}}
\end{figure}
\begin{figure}
\centering
\includegraphics{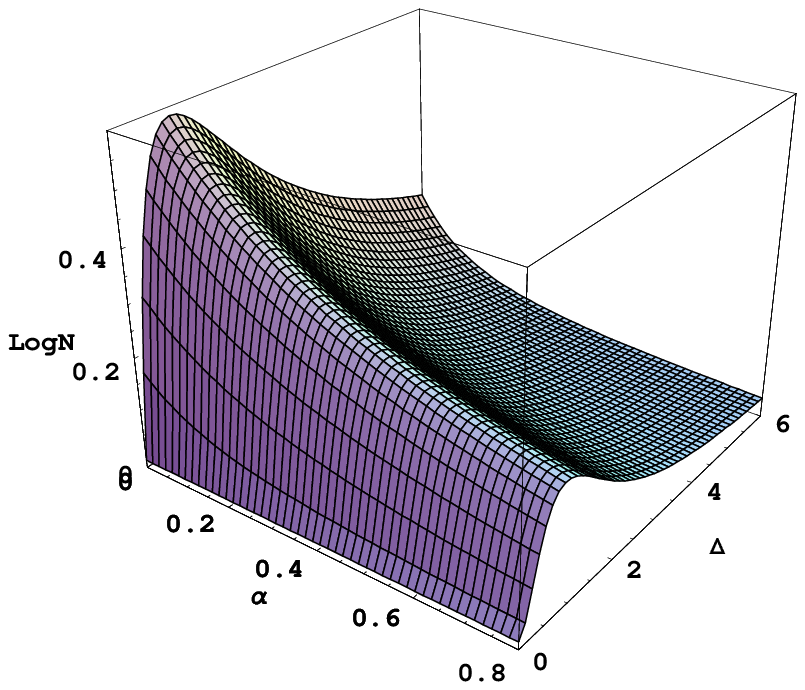}
\caption{The stationary state log-negativity $N(\rho_f)$ of the
density operator $\rho_f(\infty)$ is plotted as a function of the
mean photon number $\bar{m}_2=\alpha$ of second mode thermal field
and the detuning $\Delta$ with $g=0.5$, $\bar{m}_1=0$ and
$\theta=\frac{\pi}{2}$. \label{Fig2}}
\end{figure}
\begin{figure}
\centering
\includegraphics{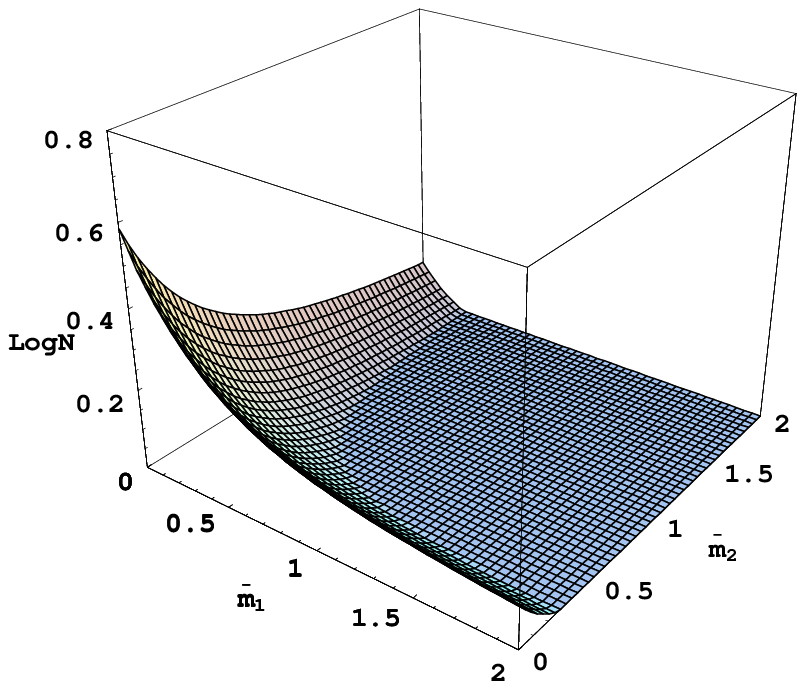}
\caption{The stationary state log-negativity $N(\rho_f)$ of the
density operator $\rho_f(\infty)$ is plotted as a function of the
mean photon number $\bar{m}_1$ and $\bar{m}_2$ of initial thermal
fields with $g=0.5$, $\Delta=1$ and $\theta=\frac{\pi}{2}$.
\label{Fig3}}
\end{figure}

In the resonant case, there is not any stationary state
entanglement between the two modes. Nevertheless, the entanglement
still arise in the forepart of the evolution, if either the
initial temperature of the thermal fields or the phase decoherence
coefficient are not too large. In Fig.4, the log-negativity
$N(\rho_f)$ of the time evolution density operator $\rho_f(t)$ is
plotted as a function of the mean photon number
$\bar{m}_1=\bar{m}_2=\alpha$ of initial thermal fields and the
time $t$. It is shown that the two-mode fields can get entangled
in the beginning of the time evolution, and become disentangled
due to the presence of decoherence. However, in the off-resonant
case, the entanglement is robust against the phase decoherence.
Fig.5 clearly displays how the two initial thermal fields get
entangled and eventually evolve into a stationary entangled state.
When the two fields are initially in thermal states, the higher
the temperature, the later the onset of entanglement between two
fields. There will not be any entanglement appearing between two
fields as their initial temperature exceeds certain threshold
value which depends on the decoherence coefficient, the coupling
strength and the detuning. Furthermore, we plot the log-negativity
as the function of the time and the mean number difference
$\delta=|\bar{m}_1-\bar{m}_2|$ with a fixed mean number sum
$\bar{m}_1+\bar{m}_2=1$. It is shown that the log-negativity
increases with the number difference at any time. In the case with
$\omega_1=\omega_2$, increasing the difference of initial
temperature of two thermal field results in enlarging the value of
$\delta$. So, one can improve the entanglement by increasing the
temperature difference in the situation that the total energy of
the initial thermal fields is fixed. The novel phenomena that
increasing the temperature difference of the thermal fields will
improve their entanglement may have some applications in the
quantum information processing, in which some subsystems are
initially in thermal equilibrium. \\
\begin{figure}
\centering
\includegraphics{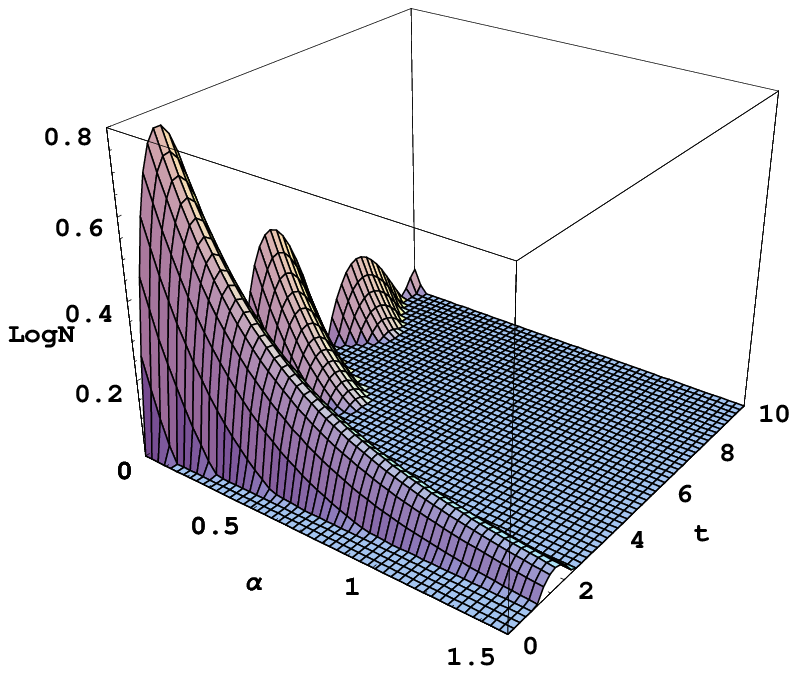}
\caption{The log-negativity $N(\rho_f)$ of the time evolution
density operator $\rho_f(t)$ is plotted as a function of the mean
photon number $\bar{m}_1=\bar{m}_2=\alpha$ of initial thermal
fields and the time $t$ with $g=0.5$, $\theta=\frac{\pi}{2}$,
$\gamma=0.5$ and $\Delta=0$. \label{Fig4}}
\end{figure}
\begin{figure}
\centering
\includegraphics{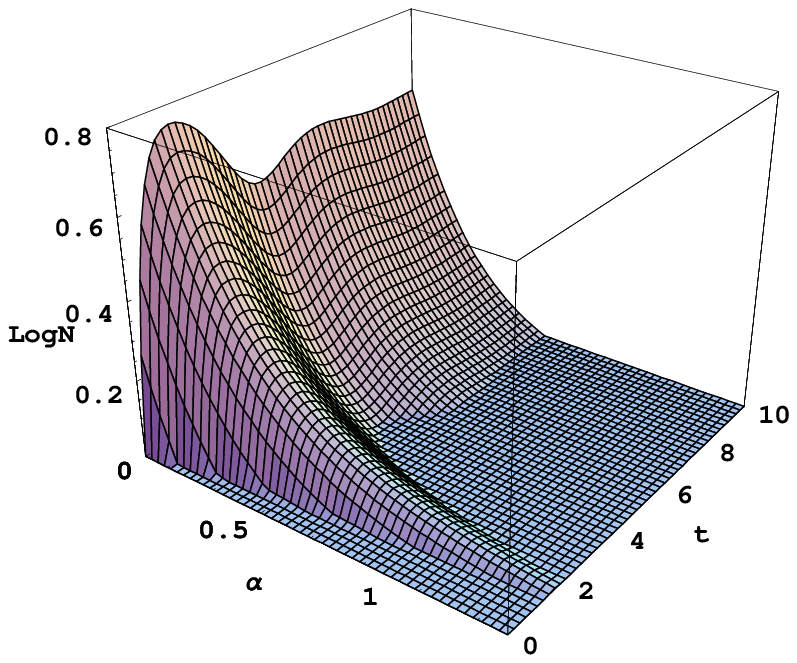}
\caption{The log-negativity $N(\rho_f)$ of the time evolution
density operator $\rho_f(t)$ is plotted as a function of the mean
photon number $\bar{m}_1=\bar{m}_2=\alpha$ of initial thermal
fields and the time $t$ with $g=0.5$, $\theta=\frac{\pi}{2}$,
$\gamma=0.5$ and $\Delta=1$. \label{Fig5}}
\end{figure}
\begin{figure}
\centering
\includegraphics{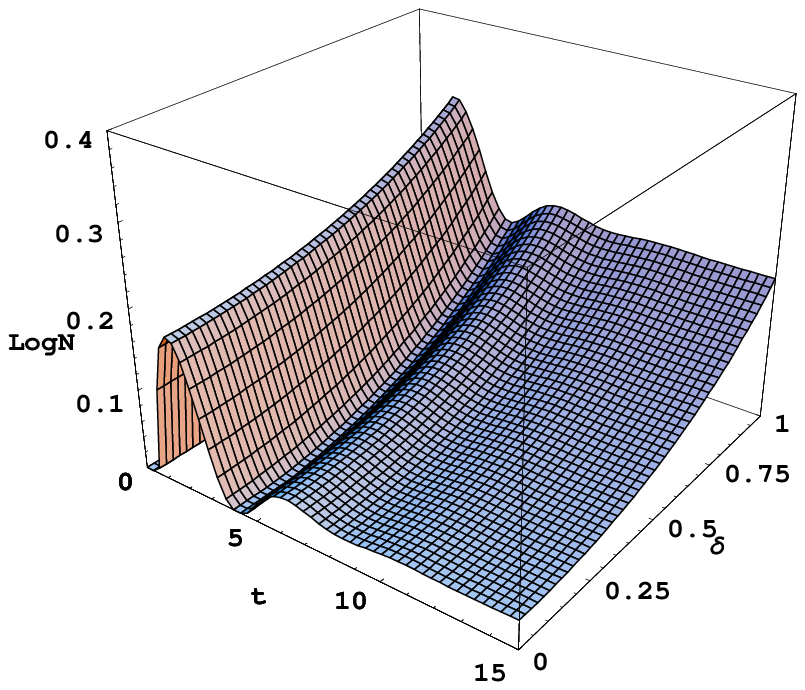}
\caption{The log-negativity $N(\rho_f)$ of the time evolution
density operator $\rho_f(t)$ is plotted as a function of the mean
photon number difference $\delta=|\bar{m}_1-\bar{m}_2|$ of initial
thermal fields and the time $t$ with $g=0.5$,
$\theta=\frac{\pi}{2}$, $\gamma=0.5$, $\bar{m}_1+\bar{m}_2=1$ and
$\Delta=1$. \label{Fig6}}
\end{figure}

\section * {IV. CONCLUSION}
\hspace*{8mm}In this paper, we investigate a possible scheme for
entangling two mode thermal fields through the quantum erasing
process, in which an atom is coupled with two mode fields via the
interaction governed by the two-mode two-photon Jaynes-Cummings
model. The influence of phase decoherence on the entanglement of
two mode fields is discussed. It is found that quantum erasing
process can transfer part of entanglement between the atom and
fields to two mode fields initially in the thermal states. The
entanglement achieved by fields heavily depends on their initial
temperature and the detuning. The entanglement of stationary state
is also investigated. It is interesting to study the entanglement
in a similar scheme in which the two-mode two-photon
Jaynes-Cummings model is replaced by the two-mode Raman coupling
Jaynes-Cummings model. Both schemes can be easily
realized in the two-dimensional ion trap. The details will be discussed elsewhere.\\

\section * {ACNOWLEDGMENT}
This project was supported by the National Natural Science
Foundation of China (Project NO. 10174066).

\end{document}